\documentclass[english,10pt,twocolumn,twoside]{IEEEtran}
\usepackage[T1]{fontenc}
\usepackage[latin9]{inputenc}
\usepackage{color}
\usepackage{bm}
\usepackage{amsmath}
\usepackage{amssymb}
\usepackage{graphicx}
\usepackage{wasysym}
\usepackage{colortbl}
\usepackage{cite}
\usepackage{babel}
\usepackage{subfig}
\usepackage{dirtytalk}
\usepackage{units}
\usepackage{subcaption}

\usepackage{comment}

\specialcomment{rimuovere}{}{}
\specialcomment{riscrivere}{\begingroup\color{orange}}{\endgroup}
\specialcomment{commentolungo}{\begingroup\color{gray}}{\endgroup}
\specialcomment{new}{\begingroup\color{blue}}{\endgroup}
\specialcomment{warning}{\begingroup\color{red}}{\endgroup}

\begin{document}
\title{Score-based Fading-Aware Decision Fusion}
\author{D.~Ciuonzo,\IEEEmembership{~Senior~Member,~IEEE}
\thanks{Manuscript received 11th May 2025; revised 18th July 2025. \\
D. Ciuonzo is with University of Naples ``Federico II'', DIETI,
Via Claudio 21, 80125 Naples, Italy. (e-mail: domenico.ciuonzo@unina.it)}
}

\markboth{IEEE Signal Processing Letters,~Vol.~*, No.~*, Month~yyyy}%
{Ciuonzo: Score-based Fading-Aware Decision Fusion}

\maketitle

\begin{abstract}
Distributed detection of an unknown deterministic signal is studied in a wireless sensor network with low-cost nodes.
Sensors apply one-bit quantization to noisy observations and transmit over Rayleigh fading to a Fusion Center (FC).
Score tests, including variants using observed Fisher information, are proposed as low-complexity alternatives to the generalized likelihood ratio test (GLRT).
The FC performs joint decoding and fusion, optimizing quantizers in a channel-aware manner to improve asymptotic detection performance. 
Simulations confirm the appeal of score-based tests under realistic wireless conditions.
\end{abstract}


\section{Introduction}
\IEEEPARstart{D}{istributed} Detection (DD)
is key in Wireless Sensor Networks (WSNs) and Internet of Things (IoT) applications~\cite{ITU2012}. Due to tight bandwidth and energy limits, sensors typically send one-bit decisions to a Fusion Center (FC). In principle, local likelihood ratio tests with one-bit quantization are optimal under both Bayesian and Neyman-Pearson criteria~\cite{Viswanathan1997}. In practice, their use is limited by \emph{exponential threshold design complexity} and \emph{incomplete sensing model knowledge}.

Fusion at the FC under the latter assumption becomes a \emph{composite hypothesis testing} problem~\cite{Kay1998}. The Generalized Likelihood Ratio Test (GLRT) is widely adopted in this context~\cite{Fang2013,gao2014quantizer,niu2023designing}, but it incurs \emph{significant computational overhead} due to the need for parameter estimation. \emph{Score tests}~\cite{Kay1998} provide an attractive alternative: they avoid Maximum Likelihood (ML) estimation under $\mathcal{H}_1$ hypothesis while being asymptotically equivalent to the GLRT.
Hence, several works have employed score tests in distributed detection, including one-bit~\cite{Ciuonzo2013b,ciuonzo2021IotJ} and multi-bit~\cite{cheng2019multibit,yang2023hybrid,mao2024multi} quantization setups, often under \emph{idealized reporting models}. 
Indeed, these approaches abstract away the physical layer and do not exploit the actual received signals or channel characteristics, namely they follow a \emph{Decode-then-Fuse} (DtF) approach~\cite{Ciuonzo2012}.
In detail, the impact of the reporting channels is typically either ($a$) neglected (e.g.~\cite{Fang2013}) or
($b$) the bit decoding phase is separated from the fusion stage (e.g.~\cite{ciuonzo2021IotJ}), resulting in a \emph{sub-optimal process}.
Conversely, \emph{Fading Aware} (FA) rules are derived by explicitly modeling the reporting channels and employing a \emph{decode-and-fuse} strategy~\cite{Ciuonzo2012} at the FC. Sadly, current FA proposals are restricted to a fully-specified sensing model, e.g.~\cite{Chen2004, al2019decision}.

Building on earlier results%
\footnote{In detail, previous work: ($a$) focused solely on the \texttt{RB} scenario, ($b$) did not design score tests based on sample-based Fisher Information, ($c$) overlooked optimal threshold design, ($d$) offered no comparison with DtF baselines.}
in \cite{ciuonzo2020SIRS}, this work addresses DD of an unknown deterministic signal in Gaussian noise over parallel-access Rayleigh fading channels.
Unlike prior works, I consider \emph{two pratical wireless reporting scenarios}:
($i$) raw quantization with Binary Phase-Shift Keying (BPSK) modulation under coherent fading~\cite{Chen2004}, proposing a Rao test;
($ii$) square quantization with On-Off Keying (OOK) modulation under statistical fading~\cite{Berger2009}, adopting a Locally-Optimum Detector (LOD).
In both cases, I also develop novel fusion rules using the \emph{Observed} Fisher Information (OFI)~\cite{deMaio2018adaptive,stoica2004model}, yielding OFI-Rao and OFI-LOD tests.
The resulting rules are tractable and avoid parameter estimation. Simulations compare them to the GLRT in a relevant WSN setting.

\noindent
\textbf{Paper organization:} Sec.~\ref{sec:System-Model} sets up the system model. Sec.~\ref{sec: GLR_GLOD} develops GLR and score tests for the FA case. Sec.~\ref{sec:Quantizer_design} designs asymptotically optimal quantizers. Sec.~\ref{sec: Simulation Results and Discussion} presents numerical results, key insights, and directions for future work.%
\footnote{\textbf{Notation:} Bold lowercase letters denote vectors; $a_{n}$ is the $n$th entry of $\bm{a}$.
$\mathbb{E}\{\cdot\}$, $(\cdot)^{T}$, $(\cdot)^{*}$, and $\Re\left(\cdot\right)$ denote expectation, transpose, conjugate, and real part. 
$u(\cdot)$ is the unit step. 
$p(\cdot)$ and $P(\cdot)$ denote probability density functions (pdf) and probability mass functions (pmf). 
$\mathcal{N}(\mu,\sigma^{2})$ and $\mathcal{N}_{\mathbb{C}}(\mu,\sigma^{2})$ are real and complex Gaussian random variables. 
$\chi_{k}^{2}\;$ and $\chi_{k}^{2}(\delta)$ are central and non-central chi-square distributions with $k$ degrees of freedom.
$\mathcal{Q}(\cdot)$
(resp. $p_{\mathcal{N}}(\cdot)$) denotes the complementary cumulative
distribution function (resp. the pdf) of a normal random variable
in its standard form, i.e. $\mathcal{N}(0,1)$;
$p_{\mathcal{N}}(\cdot;\mu,\sigma)$ denotes a Gaussian pdf, while $p_{\mathrm{exp}}(\cdot;\sigma^2)$ an exponential pdf with scale $\sigma^2$.
$\sim$ and  $\overset{a}{\sim}$ mean \say{distributed as} and \say{asymptotically distributed as}.}

\section{System Model\label{sec:System-Model}}
Consider a network of sensors $k\in\mathcal{K}\triangleq\{1,\ldots,K\}$ tasked with detecting presence/absence of an unknown deterministic scalar $\theta\in\mathbb{R}$, via a \emph{composite} binary hypothesis test~\cite{Fang2013,cheng2019multibit}:
\begin{gather}
\begin{cases}
\mathcal{H}_{0}\quad:\quad & z_{k}=n_{k},\\
\mathcal{H}_{1}\quad:\quad & z_{k}=g_{k}\,\theta+n_{k},\qquad k\in\mathcal{K};
\end{cases}\label{eq:binary_test}
\end{gather}
where $z_{k}\in\mathbb{R}$ is the sensor measurement, $g_{k}\in\mathbb{R}$ is a known gain, and $n_{k} \sim \mathcal{N}(0,\sigma_{n,k}^2)$, independent across $k$. This is a \emph{two-sided} test~\cite{Kay1998} with $\theta_0 = 0$.

Each sensor quantizes $z_{k}$ using a (tunable) threshold ($\tau_{k}$ or $\gamma_{k}$) into one bit, under (typical) strict IoT energy and bandwidth constraints~\cite{al2019decision}, and transmits it over independent flat-fading links to a FC.
When using BPSK (\( \{-1,+1\} \)), \emph{Raw Quantizers} (RQ) are adopted,
yielding $x_{k}\triangleq2\,u\,(z_{k}-\tau_{k})-1$. Differently, when using OOK (\( \{0,1\} \)),
\emph{Square-based Quantizers} (SQ) are investigated, i.e. $x_{k}\triangleq u(z_{k}^{2}-\gamma_{k})$.
For RQ, the \say{$+1$} probability under $\mathcal{H}_{1}$ equals $\alpha_{k}(\theta)\triangleq F_{n_{k}}(\tau_{k}-g_{k}\theta)$, while for $\mathcal{H}_{0}$ it is $\alpha_{0,k}=F_{n_{k}}(\tau_{k})$, with $F_{n_{k}}(\cdot)=\mathcal{Q}(\cdot\,/\sigma_{n,k})$ denoting the complementary cumulative distribution function of $n_{k}$.
Differently, for SQ, the \say{$1$} probability under $\mathcal{H}_{1}$ equals $\beta_{k}(P_{\theta})\,\triangleq\,F_{n_{k}}\left(\sqrt{\gamma_{k}}-|g_{k}|\,\sqrt{P_{\theta}}\right)+F_{n_{k}}\left(\sqrt{\gamma_{k}}+|g_{k}|\,\sqrt{P_{\theta}}\right)$, where $P_{\theta}\triangleq\theta^{2}$,
while for $\mathcal{H}_{0}$ it is $\beta_{0,k}\triangleq2\,F_{n_{k}}\left(\sqrt{\gamma_{k}}\right)$.
Under \emph{Rayleigh fading}, the received signal at the FC is:
\begin{equation}
y_{k}=h_{k}\,x_{k}+w_{k},
\end{equation}
where $h_{k}\sim\mathcal{N}_{\mathbb{C}}(0,\sigma_{h,k}^{2})$ and $w_{k}\sim\mathcal{N}_{\mathbb{C}}(0,\sigma_{w,k}^{2})$.
\emph{Perfect} Channel State Information (CSI) is assumed at the FC under BPSK~\cite{Chen2004}, and \emph{statistical CSI} under OOK~\cite{Berger2009}.
For brevity, these two scenarios are termed \texttt{RB} and \texttt{SO}, respectively.%
\footnote{\texttt{RB} denotes raw quantization with BPSK and perfect CSI, while \texttt{SO} refers to squared quantization with OOK and statistical CSI.}

Stacking received signals as $\bm{y}\triangleq\left[\begin{array}{ccc}
y_{1} & \cdots & y_{K}\end{array}\right]^{T}$, the goal is to design a \emph{computationally simple test} $\Lambda(\bm{y})\gtrless_{\mathcal{H}_{0}}^{\mathcal{H}_{1}}\gamma_{\mathrm{fc}}$, 
along with optimized thresholds ($\tau_k$'s or $\gamma_k$'s), to maximize FC detection rate $P_{\mathrm{D}}\triangleq\Pr\{\Lambda>\gamma_{\mathrm{fc}}|\mathcal{H}_{1}\}$, under a constraint on false alarm $P_{\mathrm{F}}\triangleq\Pr\{\Lambda>\gamma_{\mathrm{fc}}|\mathcal{H}_{0}\}$.
Note that, under RQ (\texttt{RB}), the test remains \emph{two-sided}. For SQ (\texttt{SO}), however, $\ensuremath{\beta_{k}(P_{\theta})}$ depends on \( P_\theta = \theta^2 \) only, leading to a \emph{one-sided} test in \( P_\theta > 0 \). 
Next section presents the fusion rule design for both cases.

\section{Score-based Fading-Aware Fusion\label{sec: GLR_GLOD}}
For \texttt{RB} case, the log-likelihood simplifies to $\ln\left[p_{\mathrm{b}}(\bm{y};\theta)\right]=\sum_{k=1}^{K}\ln\left[p_{\mathrm{b}}(y_{k};\theta)\right]$ by independence, i.e.
\begin{gather}
\ln\left[p_{\mathrm{b}}(\bm{y};\theta)\right]=\sum_{k=1}^{K}\ln\left[\mathcal{N}_{\mathbb{C}}(y_{k};h_{k},\sigma_{w,k}^{2})\,\alpha_{k}(\theta)+\right.\nonumber \\
\left.\hfill\qquad\mathcal{N}_{\mathbb{C}}(y_{k};-h_{k},\sigma_{w,k}^{2})(1-\alpha_{k}(\theta))\right]
\end{gather}
leveraging perfect CSI (known $h_k$). Similarly, in the \texttt{SO} case, the log-pdf becomes $\ln\left[p_{\mathrm{o}}(\bm{y};\theta)\right]=\sum_{k=1}^{K}\ln\left[p_{\mathrm{o}}(y_{k};\theta)\right]$, i.e.
\begin{gather}
\ln\left[p_{\mathrm{o}}(\bm{y};P_{\theta})\right]=\sum_{k=1}^{K}\ln\left[\mathcal{N}_{\mathbb{C}}(y_{k};0,\sigma_{1}^{2})\,\beta_{k}(P_{\theta})+\right.\nonumber \\
\left.\hfill\qquad\mathcal{N}_{\mathbb{C}}(y_{k};0,\sigma_{0}^{2})(1-\beta_{k}(P_{\theta}))\right]\label{eq: loglik_OOK}
\end{gather}
assuming statistical CSI with $h_{k}\sim\mathcal{N}_{\mathbb{C}}(0,\sigma_{h,k}^{2})$, and using $\sigma_{1}^{2}\triangleq(\sigma_{w,k}^{2}+\sigma_{h,k}^{2})$ , $\sigma_{0}^{2}\triangleq\sigma_{w,k}^{2}$ for brevity.
In what follows, subscripts \say{$\mathrm{b}$} and \say{$\mathrm{o}$} in the mathematical expressions refer to the \texttt{RB} and \texttt{SO} scenarios, respectively.

\noindent 
\textbf{GLRT:} A common tool for a detector in composite hypothesis
testing problems is given by the GLRT \cite{Kay1998}, which for the \texttt{RB} scenario is expressed in implicit form as:
\begin{equation}
\Lambda_{\mathrm{G}}\triangleq2\cdot\ln\left[p_{\mathrm{b}}(\bm{y};\hat{\theta}_{1})\,/\,p_{\mathrm{b}}(\bm{y};\theta_{0})\right]\,;\label{eq:GLRT_general}
\end{equation}
Here, $\hat{\theta}_1$ is the ML estimate under $\mathcal{H}_1$, i.e., $\hat{\theta}_1 \triangleq \arg\max_{\theta} p_{\mathrm{b}}(\bm{y}; \theta)$.
The same structure holds for \texttt{SO}: use the log-likelihood $p_{\mathrm{o}}(\bm{y}; P_{\theta})$ in Eq.~\eqref{eq: loglik_OOK}, and replace accordingly. In this case, the ML estimate of $P_{\theta}$ must be computed.
The maximization in Eq.~\eqref{eq:GLRT_general} shows that computing $\Lambda_{\mathrm{G}}$ involves solving an optimization problem, which \emph{adds computational cost}-e.g., via grid search or local optimization.%
\footnote{In the case of grid-search the GLRT complexity scales with $\mathcal{O}(KN_\theta)$, $N_{\theta}$ being the number of bins used to discretize the search space of $\theta$.}

\noindent
\textbf{Score Tests~\cite{Kay1998}:} To bypass GLRT difficulties, score-based tests are adopted: a Rao test for the two-sided \texttt{RB} scenario, and a LOD test for \texttt{SO}. Decision statistics are given implicitly as:
\begin{gather}
\Lambda_{\mathrm{R}}\triangleq\frac{\left(\left.\frac{\partial\ln p_{\mathrm{b}}(\bm{y};\theta)}{\partial\theta}\right|_{\theta=\theta_{0}}\right)^{2}}{\mathrm{I}_{\mathrm{b}}(\theta_{0})},\quad\Lambda_{\mathrm{L}}\triangleq\frac{\left.\frac{\partial\ln p_{\mathrm{o}}(\bm{y};P_{\theta})}{\partial P_{\theta}}\right|_{P_{\theta}=P_{\theta_{0}}}}{\sqrt{\mathrm{I}_{\mathrm{o}}(P_{\theta_{0}})}}\label{eq: LOD+Rao}
\end{gather}
Here, $\mathrm{I}_{\mathrm{b}}(\theta_0)$ and $\mathrm{I}_{\mathrm{o}}(P_{\theta_0})$ are the Fisher information (FI), defined as
$\mathrm{I}_{\mathrm{b}}(\theta)\triangleq\mathbb{E}\{\left(\partial\ln\left[p_{\mathrm{b}}(\bm{y};\theta)\right]/\partial\theta\right)^{2}\}$ and $\mathrm{I}_{\mathrm{o}}(P_{\theta})\triangleq\mathbb{E}\{\left(\partial\ln\left[p_{\mathrm{o}}(\bm{y};P_{\theta})\right]/\partial P_{\theta}\right)^{2}\}$, both evaluated at the null.
This choice is motivated by: ($i$) simplicity--avoiding $\hat{\theta}_1$ and $\hat{P}_{\theta_1}$ computation (Eq.~\eqref{eq: LOD+Rao}), yielding $\mathcal{O}(K)$ complexity if the FI is pre-computed; ($ii$) asymptotic equivalence to GLRT in the weak-signal regime~\cite{Kay1998}.
The closed-form of $\Lambda_{\mathrm{R}}$ (resp. $\Lambda_{\mathrm{L}}$) directly follows from the score and FI expressions, evaluated at $\theta = \theta_0$ (resp. $P_{\theta} = P_{\theta_0}$).
The score functions are given as:
\begin{align}
\left.\frac{\partial\ln p_{\mathrm{b}}(\bm{y};\theta)}{\,\partial\theta}\right|_{\theta=\theta_{0}}= & \sum_{k=1}^{K}\frac{s_{\mathrm{b},k}\,[\varrho_{k}(y_{k})-1]}{\varrho_{k}(y_{k})\alpha_{0,k}+[1-\alpha_{0,k}]}\label{eq:score_function_BPSK}\\
\left.\frac{\partial\ln p_{\mathrm{o}}(\bm{y};P_{\theta})}{\,\partial P_{\theta}}\right|_{P_{\theta}=P_{\theta_{0}}}= & \sum_{k=1}^{K}\frac{s_{\mathrm{o},k}\,[\varrho_{k}(y_{k})-1]}{\varrho_{k}(y_{k})\,\beta_{0,k}+[1-\beta_{0,k}]}\label{eq:score_function_OOK}
\end{align}
where $s_{\mathrm{b},k}\triangleq g_{k}p_{n_{k}}(\tau_{k})$,  $s_{\mathrm{o},k}\triangleq(g_{k}^{2}/\sigma_{n,k}^{2})\sqrt{\gamma_{k}}p_{n_{k}}(\sqrt{\gamma_{k}})$, and
\begin{gather}
\varrho_{k}(y_{k})\triangleq\begin{cases}
\exp\left[\frac{4\cdot\Re(h_{k}^{*}y_{k})}{\sigma_{w,k}^{2}}\right] & (\texttt{RB})\\
\frac{\sigma_{w,k}^{2}}{\sigma_{h,k}^{2}}\,\exp\left[\frac{|y_{k}|^{2}}{\sigma_{w,k}^{2}}\cdot\frac{\sigma_{h,k}^{2}}{\sigma_{h,k}^{2}+\sigma_{w,k}^{2}}\right] & (\texttt{SO})
\end{cases}
\end{gather}
Here $p_{n_{k}}(\cdot)=p_{\mathcal{N}}(\cdot;0,\sigma_{n,k}^{2})$ denotes the pdf of $n_k$. For the \texttt{RB} case, $\mathrm{I}_{\mathrm{b}}(\theta_0)$ admits the explicit expression
\begin{gather}
\mathrm{I}_{\mathrm{b}}(\theta_{0})=\sum_{k=1}^{K}s_{\mathrm{b},k}^{2}\cdot\{\alpha_{0,k}\,\phi_{k}^{+}(\tau_{k})+\left[1-\alpha_{0,k}\right]\phi_{k}^{-}(\tau_{k})\}\label{eq: FI_BPSK}
\end{gather}
and a similar form holds for the \texttt{SO} case:
\begin{gather}
\mathrm{I}_{\mathrm{o}}(P_{\theta_{0}})=\sum_{k=1}^{K}s_{\mathrm{o},k}^{2}\cdot\{\beta_{0,k}\,\varphi_{k}^{1}(\gamma_{k})+[1-\beta_{0.k}]\,\varphi_{k}^{0}(\gamma_{k})\}\label{eq: FI_OOK}
\end{gather}
Definitions of $\phi_k^{\pm}(\tau_k)$ and $\varphi_{k}^{i}(\gamma_{k})$ are given in Eqs.~~\eqref{eq: FI_term_BPSK} and \eqref{eq: FI_term_OOK}, respectively, at the top of the next page.
In the above definitions the auxiliary terms $\Xi_k \triangleq 2|h_k|^2 / \sigma_{w,k}^2$ and $\overline{\Xi}_k \triangleq \sigma_{h,k}^2 / \sigma_{w,k}^2$ are used.
Derivations are omitted for brevity and can be obtained similarly as~\cite{ciuonzo2020SIRS}.
Combining Eqs.~(\ref{eq:score_function_BPSK}-\ref{eq:score_function_OOK}), (\ref{eq: FI_BPSK}-\ref{eq: FI_OOK})~and~(\ref{eq: FI_term_BPSK}-\ref{eq: FI_term_OOK}),
$\Lambda_{\mathrm{R}}$ and 
$\Lambda_{\mathrm{L}}$ are obtained in \emph{closed form}.
\begin{figure*}
\begin{align}
\phi_{k}^{\pm}(\tau_{k})\triangleq & \int_{-\infty}^{+\infty}d\xi_{k}\left(\frac{\exp(2\,\xi_{k})-1}{\exp(2\,\xi_{k})\,F_{n_{k}}(\tau_{k})+\left(1-F_{n_{k}}(\tau_{k})\right)}\right)^{2}p_{\mathcal{N}}\left(\xi_{k};\pm\Xi_{k},\Xi_{k}\right)\label{eq: FI_term_BPSK}\\
\varphi_{k}^{i}(\gamma_{k})\triangleq & \int_{-\infty}^{+\infty}d\varepsilon_{k}\left(\frac{\frac{1}{1+\overline{\Xi}_{k}}\,\exp(\frac{\varepsilon_{k}}{\sigma_{w,k}^{2}}\cdot\frac{\overline{\Xi}_{k}}{1+\overline{\Xi}_{k}})-1}{\frac{1}{1+\overline{\Xi}_{k}}\,\,\exp(\frac{\varepsilon_{k}}{\sigma_{w,k}^{2}}\cdot\frac{\overline{\Xi}_{k}}{1+\overline{\Xi}_{k}})\,2F_{n_{k}}(\sqrt{\gamma_{k}})+\left(1-2F_{n_{k}}(\sqrt{\gamma_{k}})\right)}\right)^{2}p_{\mathrm{exp}}\left(\varepsilon_{k};\sigma_{i}^{2}\right)\label{eq: FI_term_OOK}
\end{align}
\hrulefill 
\vspace*{-2pt}
\end{figure*}

\begin{figure*}
\subfloat[Objective for design of $\tilde{\tau}_{k}^{\star}$ (viz. $\tau_{k}^{\star}$) in \texttt{RB} case (by varying $\Xi_k$).\label{fig:RQ objective}]{\includegraphics[width=0.42\paperwidth]{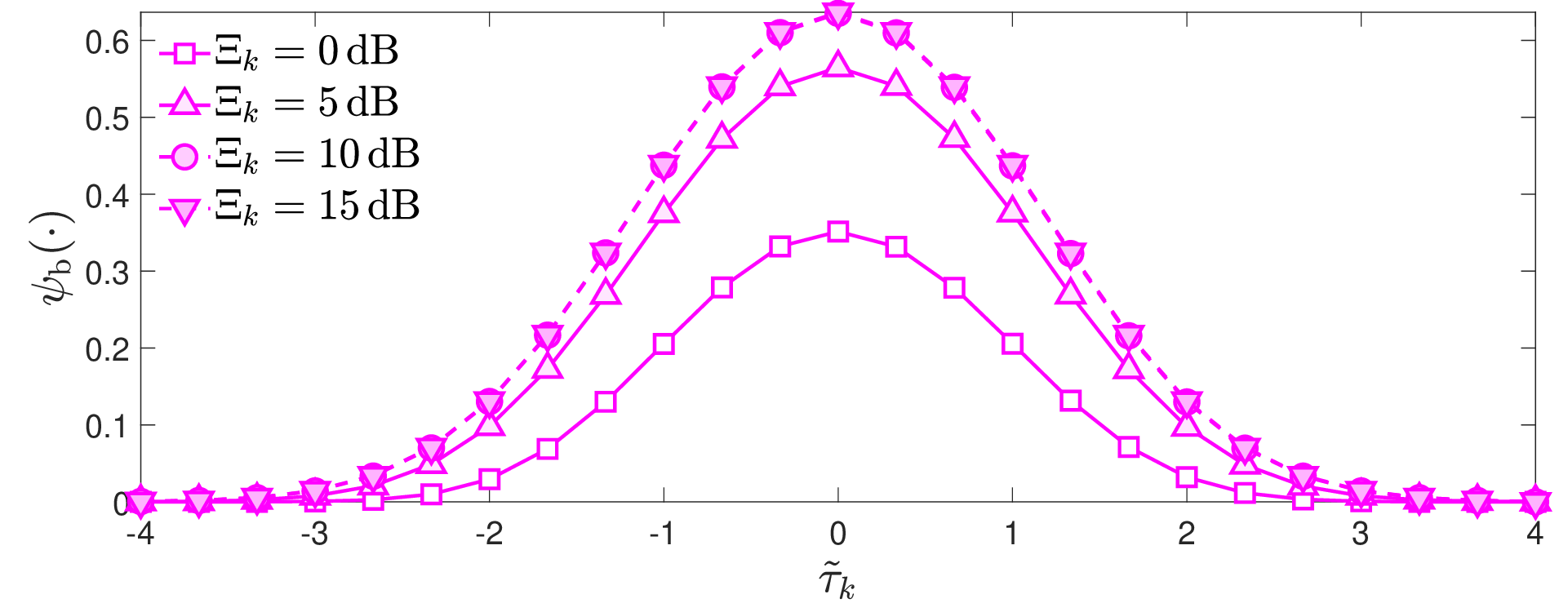}}\subfloat[Objective for design of $\beta_{0,k}^{\star}$ (viz. $\gamma_{k}^{\star}$)
in \texttt{SO} case (by varying $\overline{\Xi}_k$).\label{fig:SQ objective}]{\includegraphics[width=0.42\paperwidth]{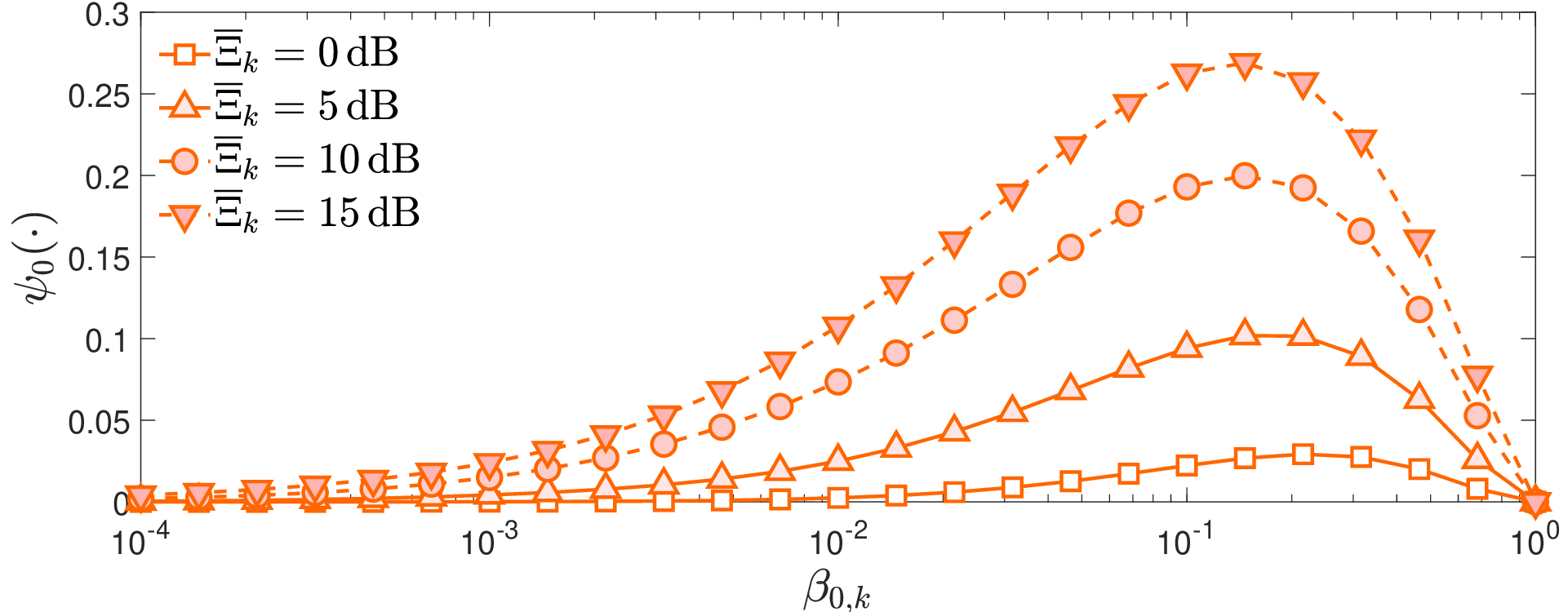}}\caption{Threshold objectives for \texttt{RB} (subfig. a), and \texttt{SO} (subfig. b), shown for different channel SNR values.
}
\end{figure*}

\noindent
\textbf{OFI-Score Tests~\cite{deMaio2018adaptive}:} Direct evaluation of the FI expressions in Eqs.~\eqref{eq: FI_BPSK} and \eqref{eq: FI_OOK} is nontrivial, as they hinge on integral terms in Eqs.~\eqref{eq: FI_term_BPSK} and \eqref{eq: FI_term_OOK}, which in turn depend on $\Xi_k$ and $\overline{\Xi}_k$ ($\propto$ to the instantaneous and average channel SNR).
To bypass this, \emph{alternative score-based statistics} are proposed: 
\begin{equation}
\Lambda_{\mathrm{OR}}\triangleq\frac{\left(\left.\frac{\partial\ln p_{\mathrm{b}}(\bm{y};\theta)}{\partial\theta}\right|_{\theta=\theta_{0}}\right)^{2}}{\left.-\frac{\partial^{2}\ln p_{\mathrm{b}}(\bm{y};\theta)}{\partial\theta^{2}}\right|_{\theta=\theta_{0}}}\;\Lambda_{\mathrm{OL}}\triangleq\frac{\left.\frac{\partial\ln p_{\mathrm{o}}(\bm{y};P_{\theta})}{\partial P_{\theta}}\right|_{P_{\theta}=P_{\theta_{0}}}}{\sqrt{\left.-\frac{\partial^{2}\ln p_{\mathrm{o}}(\bm{y};P_{\theta})}{\partial P_{\theta}^{2}}\right|_{P_{\theta}=P_{\theta_{0}}}}}\label{eq:OFI_score_general}
\end{equation}
Here, the FI is replaced by its \emph{sample-based estimate} (OFI)~\cite{stoica2004model}, evaluated at $\theta_0$ or $P_{\theta_0}$.
After simplification%
\footnote{Calculation of second derivative of the log-likelihood in Eq.~\eqref{eq:OFI_score_general} is greatly simplified by noting that $\lim_{\theta\rightarrow\theta_{0}}\partial^{2}\alpha_{k}(\theta)\,/\,\partial\theta^{2}=0$
and $\lim_{P_{\theta}\rightarrow P_{\theta_{0}}}\partial^{2}\beta_{k}(P_{\theta})\,/\,\partial P_{\theta}^{2}=0$.
This leads to the following equivalences:
\begin{gather}
\ensuremath{\left.-\partial^{2}\ln p_{\mathrm{b}}(\bm{y};\theta)/\,\partial\theta^{2}\right|_{\theta=\theta_{0}}=}\sum_{k=1}^{K}(\left.\partial\ln p_{\mathrm{b}}(y_{k};\theta)\,/\,\partial\theta\right|_{\theta=\theta_{0}})^{2}\\
\left.-\partial^{2}\ln p_{\mathrm{o}}(\bm{y};P_{\theta})/\partial P_{\theta}^{2}\right|_{P_{\theta}=P_{\theta_{0}}}=\sum_{k=1}^{K}(\left.\partial\ln p_{\mathrm{o}}(y_{k};P_{\theta})/\partial P_{\theta}\right|_{P_{\theta}=P_{\theta_{0}}})^{2}\nonumber 
\end{gather}
}, %
the explicit form of OFI-Rao and OFI-LOD rules is obtained as:
\begin{align}
\Lambda_{\mathrm{OR}}= & {\displaystyle \frac{\left\{ \sum_{k=1}^{K}\frac{s_{\mathrm{b},k}\cdot[\varrho_{k}(y_{k})-1]}{\varrho_{k}(y_{k})\,\alpha_{0,k}+[1-\alpha_{0,k}]}\right\} ^{2}}{\sum_{k=1}^{K}\frac{s_{\mathrm{b},k}^{2}\cdot[\varrho_{k}(y_{k})-1]^{2}}{\left\{ \varrho_{k}(y_{k})\,\alpha_{0,k}+[1-\alpha_{0,k}]\right\} ^{2}}}}\label{eq:OFI-LOD Test explicit- CA case}\\
\Lambda_{\mathrm{OL}}= & {\displaystyle \frac{\sum_{k=1}^{K}\frac{s_{\mathrm{o},k}\cdot[\varrho_{k}(y_{k})-1]}{\varrho_{k}(y_{k})\,\beta_{0,k}+[1-\beta_{0,k}]}}{\sqrt{\sum_{k=1}^{K}\frac{s_{\mathrm{o},k}^{2}\cdot[\varrho_{k}(y_{k})-1]^{2}}{\left\{ \varrho_{k}(y_{k})\,\beta_{0,k}+[1-\beta_{0,k}]\right\} ^{2}}}}}\label{eq:OFI-Rao Test explicit- CA case}
\end{align}
Fusion rules for \texttt{RB} ($\Lambda_{\mathrm{R}}, \Lambda_{\mathrm{OR}}, \Lambda_{\mathrm{G}}$) depend on thresholds $\tau_k$'s; likewise, \texttt{SO} rules ($\Lambda_{\mathrm{L}}, \Lambda_{\mathrm{OL}}, \Lambda_{\mathrm{G}}$) depend on $\gamma_k$'s.
Threshold vectors $\bm{\tau} = [\tau_1, \ldots, \tau_K]^T$ and $\bm{\gamma} = [\gamma_1, \ldots, \gamma_K]^T$ are \emph{design parameters}, with their optimization objectives derived next.

\section{Asymptotically-Optimal Quantizer design\label{sec:Quantizer_design}}
We recall that Rao and LOD test statistics $\Lambda_{\mathrm{R}}$/$\Lambda_{\mathrm{OR}}$ and $\Lambda_{\mathrm{L}}$/$\Lambda_{\mathrm{OL}}$ (as well as $\Lambda_{\mathrm{G}}$ in both cases) are distributed (under an asymptotic, weak-signal, assumption) as~\cite{Kay1998,deMaio2018adaptive}:
\begin{eqnarray}
\Lambda_{\mathrm{(\mathrm{O)}R}}\overset{a}{\sim} & \chi_{1}^{2}\;/\;\chi_{1}^{2}(\lambda_{Q}(\bm{\tau})) & [\mathcal{H}_{0}/\mathcal{H}_{1}]\\
\Lambda_{\mathrm{(\mathrm{O)}L}}\overset{a}{\sim} & \mathcal{N}(0,1)\;/\;\mathcal{N}(\delta_{Q}(\bm{\gamma}),1) & [\mathcal{H}_{0}/\mathcal{H}_{1}]
\end{eqnarray}
where $\lambda_{Q}(\bm{\tau})\triangleq(\theta_{1}-\theta_{0})^{2}\cdot\mathrm{I}_{\mathrm{b}}(\theta_{0},\bm{\tau})$ and $\delta_{Q}(\bm{\gamma})\triangleq(P_{\theta_{1}}-P_{\theta_{0}})\cdot\sqrt{\mathrm{I}_{\mathrm{o}}(P_{\theta_{0}},\bm{\gamma})}$.
Also, $\theta_1$ and $P_{\theta_1}$ represent the true values of the
target signal and power, respectively, under $\mathcal{H}_1$.
Maximizing $\lambda_{Q}(\bm{\tau})$ and $\delta_{Q}(\bm{\gamma})$ improves detection performance.
Since both correspond to maximizing $\mathrm{I}_{\mathrm{b}}(\theta_0, \bm{\tau})$ and $\mathrm{I}_{\mathrm{o}}(P_{\theta_0}, \bm{\gamma})$, respectively, the two optimization problems are
\begin{align}
\arg\max_{\{\tau_{1},\ldots,\tau_{K}\}} & \mathrm{I}_{\mathrm{b}}(\theta_{0},\bm{\tau})\qquad\arg\max_{\{\gamma_{1},\ldots,\gamma_{K}\}}\mathrm{I}_{\mathrm{o}}(P_{\theta_{0}},\bm{\gamma})
\end{align}
These $K$-dimensional problems \emph{decouple} across sensors, yielding $K$ independent \emph{scalar problems}.
For \texttt{RB}, the objective function becomes (starting from Eq.~\eqref{eq: FI_BPSK}, using the reparametrization $\tau_{k}=\tilde{\tau}_{k}\cdot\sigma_{n,k}^{2}$ and dropping irrelevant terms):
\begin{gather}
\max_{\tilde{\tau}_{k}\in\mathbb{R}}\ \psi_{\mathrm{b}}(\tilde{\tau}_{k})\triangleq p_{\mathcal{N}}^{2}(\tilde{\tau}_{k})\times\nonumber \\{}
[\mathcal{Q}(\tilde{\tau}_{k})\,\phi_{k}^{+}(\tilde{\tau}_{k})+(1-\mathcal{Q}(\tilde{\tau}_{k}))\,\phi_{k}^{-}(\tilde{\tau}_{k})]
\end{gather}
and for \texttt{SO} (starting from Eq.~\eqref{eq: FI_OOK}, using the reparametrization $\gamma_{k}=\sigma_{n,k}^{2}(\mathcal{Q}^{-1}(\beta_{0,k}\,/\,2))^{2}$ and dropping irrelevant terms):
\begin{gather}
\max_{\beta_{0,k}\in[0,1]}\ \psi_{0}(\beta_{0,k})\triangleq p_{\mathcal{N}}^{2}(\mathcal{Q}^{-1}(\beta_{0,k}/2))\cdot[\mathcal{Q}^{-1}(\beta_{0,k}/2)]^{2}\times\nonumber \\{}
[\beta_{0,k}\,\varphi_{k}^{1}(\beta_{0,k})+(1-\beta_{0,k})\,\varphi_{k}^{0}(\beta_{0,k})\big]\label{eq: psi0_OOK}
\end{gather}
For \texttt{RB} case, the objective function is \emph{symmetric} due to the evenness of $p_{\mathcal{N}}^2(\tilde{\tau}_k)$ and since $\mathcal{Q}(-\tilde{\tau}_k) = 1 - \mathcal{Q}(\tilde{\tau}_k)$, which swaps the symmetric integrals. Also, it is \emph{unimodal}: this follows from the Gaussian pdf log-concavity and the monotonicity of the integrand, as the unimodal envelope $p_{\mathcal{N}}^2(\tilde{\tau}_k)$ preserves the unique mode at zero.
\emph{The optimum is thus attained at $\tilde{\tau}_{k}^{\star}=\tau_{k}^{\star}=0$}.
This result aligns with optimality conditions under ideal and BSC channel models~\cite{Rousseau2003,ciuonzo2021IotJ}.
In contrast, the \texttt{SO} case requires a 1-D search to obtain $\beta_{0,k}^\star$ (viz. $\gamma_k^\star$).
Representative objective function profiles for \texttt{RB}/\texttt{SO} are shown in Figs.~\ref{fig:RQ objective}-\ref{fig:SQ objective} for varying reporting channel conditions.
Substituting $\tau_k^* = 0$ further simplifies Rao and OFI-Rao as
\begin{align}
\Lambda_{\mathrm{R}}^{\star}= & \frac{\left[\sum_{k=1}^{K}g_{k}\,p_{n_{k}}(0)\,\tanh\left(2\,\Re(h_{k}^{*}y_{k})/\sigma_{w,k}^{2}\right)\right]^{2}}{\sum_{k=1}^{K}g_{k}^{2}\,p_{n_{k}}^{2}(0)\,\mathbb{E}_{\xi_{k}}\{\tanh^{2}(\xi_{k})\}}\\
\Lambda_{\mathrm{\mathrm{OR}}}^{\star}= & {\displaystyle \frac{\left(\sum_{k=1}^{K}g_{k}\,p_{n_{k}}(0)\,\tanh\left(2\,\Re(h_{k}^{*}y_{k})/\sigma_{w,k}^{2}\right)\right)}{\sum_{k=1}^{K}g_{k}^{2}\,p_{n_{k}}^{2}(0)\,\tanh^{2}\left(2\,\Re(h_{k}^{*}y_{k})/\sigma_{w,k}^{2}\right)}^{2}}
\end{align}
with $\xi_k \sim \frac{1}{2}\mathcal{N}(\Xi_k, \Xi_k) + \frac{1}{2}\mathcal{N}(-\Xi_k, \Xi_k)$.
Hence, the threshold-optimized non-centrality for \texttt{RB} equals $\lambda_{Q}^{*}=4\,\theta_{1}^{2}\sum_{k=1}^{K}g_{k}^{2}\,p_{n_{k}}^{2}(0)\,\mathbb{E}_{\xi_{k}}\{\tanh^{2}(\xi_{k})\}$.

\noindent
\textbf{Comparison with Prior Work:} Under a Binary Symmetric Channel (BSC) model, as in the DtF scheme, the optimized non-centrality is given by $\lambda_Q^\star = 4\, \theta_1^2 \sum_k g_k^2\, p_{n_k}^2(0)\, (1 - 2 P_{e,k})^2$~\cite{Fang2013,Ciuonzo2013b}. In the ideal case (no channel errors), this term reduces to $\lambda_Q^\star = 4\, \theta_1^2 \sum_k g_k^2\, p_{n_k}^2(0)$. In contrast, the impact of coherent fading is captured via $\mathbb{E}_{\xi_k}\{\tanh^2(\xi_k)\}$.

\section{Discussion of Results and Wrap-Up \label{sec: Simulation Results and Discussion}}
\noindent
\textbf{Considered Setup:} A WSN with $K = 10$ sensors is considered to assess the proposed score tests.
Sensor thresholds $\tau_{k}^{\star}$'s and $\gamma_{k}^{\star}$'s are set according to Sec.~\ref{sec:Quantizer_design}. 
The observation SNR at sensor $k$ is defined as $\mathrm{SNR}_{k}^{\mathrm{obs}}\triangleq(g_{k}^{2}\,\theta^{2})\,/\sigma_{n,k}^{2}$. 
The reporting channel SNR for \texttt{RB} case is given by $\mathrm{SNR}_k^{\mathrm{chan}} \triangleq \mathbb{E}\{|h_k|^2\}/\sigma_{w,k}^2$, while for the \texttt{SO} case, it is defined as $\mathrm{SNR}_k^{\mathrm{chan}} \triangleq (\Pr(\mathcal{H}_1)\, \mathbb{E}\{|h_k|^2\})/\sigma_{w,k}^2$, accounting for OOK energy efficiency. A prior probability of $\Pr(\mathcal{H}_1) = 0.3$ is used throughout.
A \emph{homogeneous scenario} is assumed, with $\mathrm{SNR}_k^{\mathrm{obs}} = \mathrm{SNR}^{\mathrm{obs}}$ and $\mathrm{SNR}_k^{\mathrm{chan}} = \mathrm{SNR}^{\mathrm{chan}}$ for all $k$. Each figure is based on $2\cdot10^5$ Monte Carlo runs.

\noindent
\textbf{Existing Baselines:}
In addition to the (FA) GLR in Eq.~\eqref{eq:GLRT_general}, proposed rules are compared against existing Decode-then-Fuse (DtF) schemes: DtF-GLR~\cite{Fang2013}, DtF-Rao~\cite{Ciuonzo2013b}, and DtF-LOD~\cite{Gao2015}.
DtF methods decouple bit decoding and fusion~\cite{Ciuonzo2012}: bits are decoded via ML, with error probability $P_{e,k}$.
The latter parameter is then used to model channel uncertainty via a BSC model in the fusion rule design.
Bit decisions are there denoted by $\hat{x}_{\mathrm{b},k}$ (\texttt{RB}) and $\hat{x}_{\mathrm{o},k}$ (\texttt{SO}).
The effective symbol-level probabilities are $\tilde{\alpha}_{0,k} = \alpha_{0,k}(1\!-\!P_{e,k}) + (1\!-\!\alpha_{0,k})P_{e,k}$ and
$\tilde{\beta}_{0,k} = \beta_{0,k}(1\!-\!P_{e,k}) + (1\!-\!\beta_{0,k})P_{e,k}$, yielding
$\Lambda_{{\scriptscriptstyle \mathrm{G}}}^{\mathrm{DtF}}=2\,\sum_{k=1}^{K}\ln[P_{\mathrm{b}}(\hat{x}_{\mathrm{b},k};\hat{\theta}_{1})\,/\,P_{\mathrm{b}}(\hat{x}_{\mathrm{b},k};\theta_{0})]$
(or $\Lambda_{{\scriptscriptstyle \mathrm{G}}}^{\mathrm{DtF}}=2\,\sum_{k=1}^{K}\ln[P_{\mathrm{o}}(\hat{x}_{\mathrm{o},k};\hat{P}_{\theta_{1}})\,/\,P_{\mathrm{o}}(\hat{x}_{\mathrm{o,}k};P_{\theta_{0}})]$) and 
\begin{align}
\ensuremath{\Lambda_{{\scriptscriptstyle \mathrm{R}}}^{\mathrm{DtF}}=} & \frac{\left\{ \sum_{k=1}^{K}\frac{s_{\mathrm{b},k}\,(\frac{\widehat{x}_{\mathrm{b},k}+1}{2}-\tilde{\alpha}_{0,k})\,(1-2P_{\mathrm{e},k})}{\tilde{\alpha}_{0,k}\left[1-\tilde{\alpha}_{0,k}\right]}\right\} ^{2}}{\sum_{k=1}^{K}\frac{s_{\mathrm{b},k}^{2}\,(1-2P_{\mathrm{e},k})^{2}}{\tilde{\alpha}_{0,k}\left[1-\tilde{\alpha}_{0,k}\right]}}\label{eq: DtF_Rao}\\
\Lambda_{{\scriptscriptstyle \mathrm{L}}}^{\mathrm{DtF}}= & \frac{\sum_{k=1}^{K}\frac{s_{\mathrm{o},k}\,(\widehat{x}_{\mathrm{o,}k}-\tilde{\beta}_{0,k})\,(1-2P_{\mathrm{e},k})}{\tilde{\beta}_{0,k}\left[1-\tilde{\beta}_{0,k}\right]}}{\sqrt{\sum_{k=1}^{K}\frac{s_{\mathrm{o},k}^{2}\,(1-2P_{\mathrm{e},k})^{2}}{\tilde{\beta}_{0,k}\left[1-\tilde{\beta}_{0,k}\right]}}}\label{eq: DtF_LOD}
\end{align}
Thresholds in these rules follow the corresponding DtF-based FI design: $\tau_k^\star = 0$ for raw quantization~\cite{Ciuonzo2013b}, and a 1D line search (similarly as Eq.~\eqref{eq: psi0_OOK}) for square quantization~\cite{ciuonzo2021IotJ}.

\begin{figure}[t]
\centering{}\includegraphics[width=1\columnwidth]{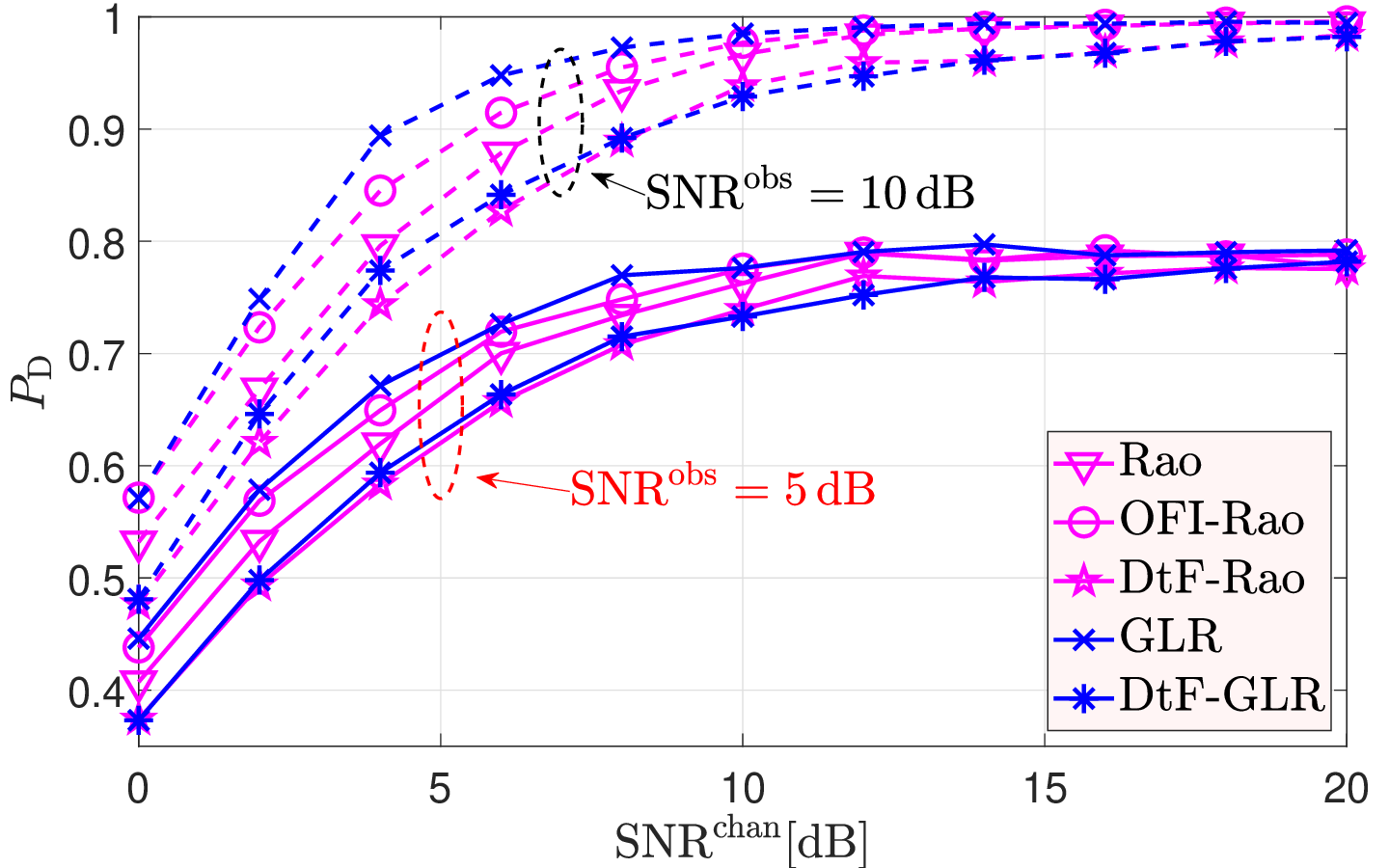}\caption{[\texttt{RB} scenario] $\ensuremath{P_{\mathrm{D}}}$ vs. $\mathrm{SNR}^{\mathrm{chan}}$ (false-alarm $P_{\mathrm{F}}=0.01$) for GLR and Rao tests. Two sensing quality levels: $\mathrm{SNR}^{\mathrm{obs}}\in\{5,10\}\,\mathrm{dB}$. Sensor thresholds set as $\tau_{k}^{\star}=0$.\label{fig: Pdo vs Pf0 (GLR vs Rao) SNR_sen}}
\end{figure}

\begin{figure}[t]
\centering{}\includegraphics[width=1\columnwidth]{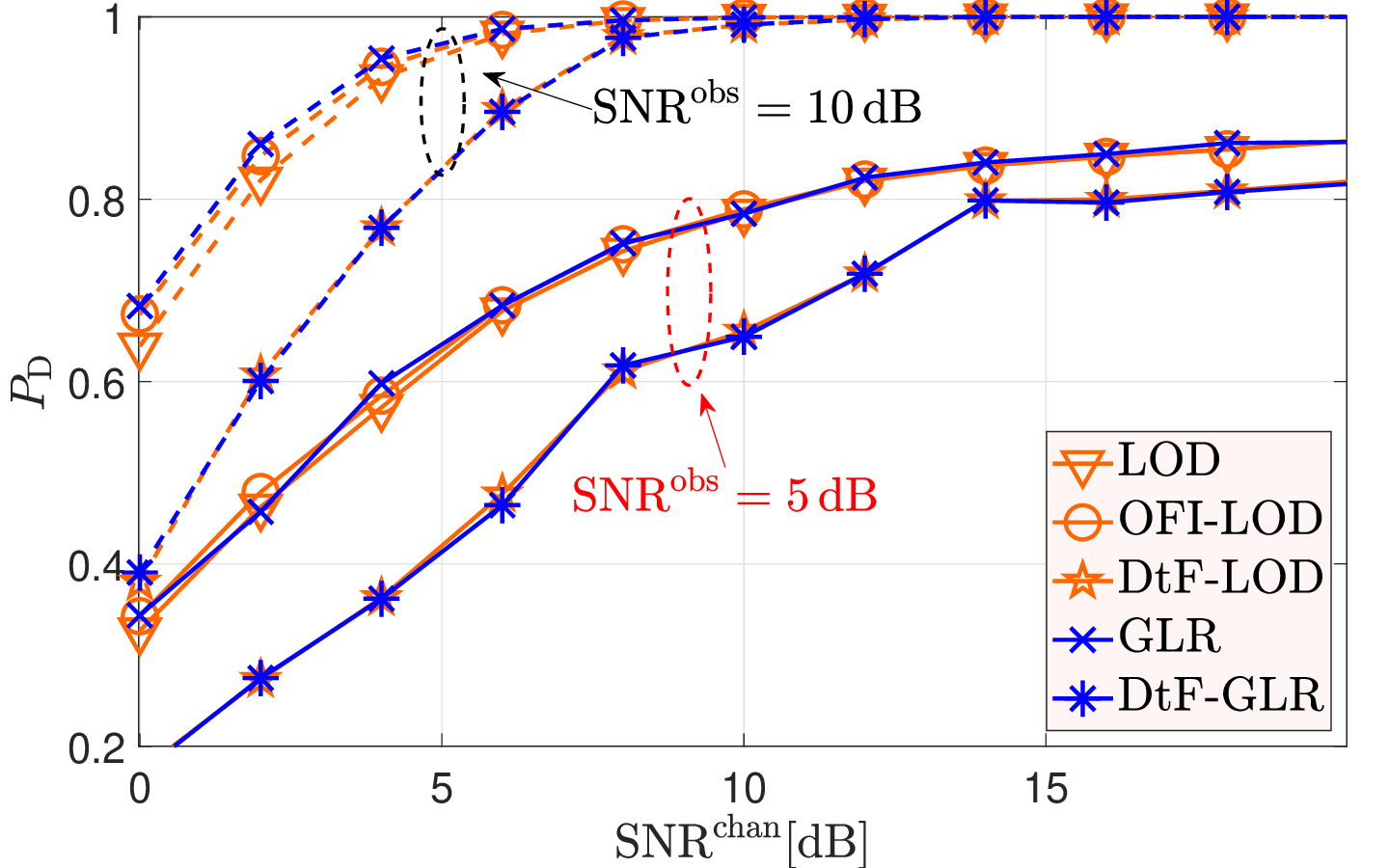}\caption{[\texttt{SO} scenario] $\ensuremath{P_{\mathrm{D}}}$ vs. $\mathrm{SNR}^{\mathrm{chan}}$ (false-alarm $P_{\mathrm{F}}=0.01$) for GLR and LOD tests. 
Two sensing quality levels:
$\mathrm{SNR}^{\mathrm{obs}}\in\{5,10\}\,\mathrm{dB}$. Sensor thresholds $\gamma_k^{\star}$ set as Eq.~\eqref{eq: psi0_OOK}.\label{fig: Pdo vs Pf0 (GLR vs LOD) SNR_chan}}
\end{figure}

\noindent
\textbf{Discussion of Results:}%
\footnote{Further analyses are provided as supplemental material to this work.}
Figs.~\ref{fig: Pdo vs Pf0 (GLR vs Rao) SNR_sen} (\texttt{RB})~and~\ref{fig: Pdo vs Pf0 (GLR vs LOD) SNR_chan} (\texttt{SO}) compare
the GLR and score tests by reporting the $\ensuremath{P_{\mathrm{D}}}$
vs. $\mathrm{SNR}^{\mathrm{chan}}$  (subject to a WSN false-alarm rate $P_{\mathrm{F}}=0.01$),
while two different cases of sensing SNR are considered:
$\mathrm{SNR}^{\mathrm{obs}}\in\{5,10\}\,\mathrm{dB}$. 
Such analysis is performed to appreciate the effect of ($a$) reporting channel quality (via $\mathrm{SNR}^{\mathrm{chan}}$) and ($b$) unknown signal strength (via $\mathrm{SNR}^{\mathrm{obs}}$) on the capabilities of the considered fusion rules. 
First, results highlight \emph{comparable
performance of the proposed FA score-tests with the FA-GLR}, despite reduced complexity, in both their exact or sample-based FI forms.
Interestingly, $\Lambda_{\mathrm{OR}}$ and $\Lambda_{\mathrm{OL}}$ \emph{slightly outperform their counterpart based on exact FI} in all the cases considered. Second, DtF-based rules (both GLR and score tests) incur in a \emph{significant gap}, especially at low-to-medium $\mathrm{SNR}^{\mathrm{chan}}$, which is an expected operating point of energy-efficient WSNs (e.g. $-16\%$ $P_{\mathrm{D}}$ loss at $\mathrm{SNR}^{\mathrm{chan}} = 4 \,\mathrm{dB}$ when $\mathrm{SNR}^{\mathrm{obs}}=10\,\mathrm{dB}$ for \texttt{SO} case). 
Finally, increasing $\mathrm{SNR}^{\mathrm{obs}}$ from $5\,\mathrm{dB}$ to $10\,\mathrm{dB}$ yields tangible gains in FA score tests performance--despite their design being rooted in a weak-signal regime~\cite{Kay1998}. 
At high $\mathrm{SNR}^{\mathrm{chan}}$, WSN detection rate $P_{\mathrm{D}}$ is \emph{limited by sensing uncertainty} when $\mathrm{SNR}^{\mathrm{obs}} = 5\,\mathrm{dB}$, while this effect is negligible at $10\,\mathrm{dB}$, allowing \emph{ideal performance}.

\noindent
\textbf{Take-aways:} this letter proposed FA score tests (and sample-based FI counterparts) as low-complexity alternatives to the GLRT for distributed detection of an unknown parameter $\theta$ under Gaussian noise, 1-bit quantization, and Rayleigh fading--explicitly modeled at the FC.
Two scenarios were considered: raw quantization with coherent BPSK (\texttt{RB}) and square quantization with blind OOK (\texttt{SO}, targeting rare-event detection).
An asymptotic threshold design was proposed and validated: $\tau_k^\star = 0$ for \texttt{RB}; 1D optimization for \texttt{SO}.
Simulations confirm that FA score tests ($a$) match FA GLRT also in finite-sensor setups and ($b$) outperform DtF-based previous proposals.
\emph{Future work} will deal with localized phenomena sensing setups~\cite{Ciuonzo2017}, imperfect channel estimates, and interference-limited scenarios~\cite{Ciuonzo2015}.

\bibliographystyle{IEEEtran}
\bibliography{IEEEabrv,sensor_networks}

\begin{thebibliography}{10}
\providecommand{\url}[1]{#1}
\csname url@samestyle\endcsname
\providecommand{\newblock}{\relax}
\providecommand{\bibinfo}[2]{#2}
\providecommand{\BIBentrySTDinterwordspacing}{\spaceskip=0pt\relax}
\providecommand{\BIBentryALTinterwordstretchfactor}{4}
\providecommand{\BIBentryALTinterwordspacing}{\spaceskip=\fontdimen2\font plus
\BIBentryALTinterwordstretchfactor\fontdimen3\font minus \fontdimen4\font\relax}
\providecommand{\BIBforeignlanguage}[2]{{%
\expandafter\ifx\csname l@#1\endcsname\relax
\typeout{** WARNING: IEEEtran.bst: No hyphenation pattern has been}%
\typeout{** loaded for the language `#1'. Using the pattern for}%
\typeout{** the default language instead.}%
\else
\language=\csname l@#1\endcsname
\fi
#2}}
\providecommand{\BIBdecl}{\relax}
\BIBdecl

\bibitem{ITU2012}
{ITU-T Rec. Y.2060 (06/2012)}, ``Overview of the {Internet} of {Things} {(IoT)},'' Jun. 2012.

\bibitem{Viswanathan1997}
R.~Viswanathan and P.~K. Varshney, ``Distributed detection with multiple sensors - {Part I: Fundamentals},'' \emph{Proc. {IEEE}}, vol.~85, no.~1, pp. 54--63, Jan. 1997.

\bibitem{Kay1998}
S.~M. Kay, \emph{Fundamentals of Statistical Signal Processing, Volume 2: Detection Theory}.\hskip 1em plus 0.5em minus 0.4em\relax Prentice Hall PTR, Jan. 1998.

\bibitem{Fang2013}
J.~Fang, Y.~Liu, H.~Li, and S.~Li, ``One-bit quantizer design for multisensor {GLRT} fusion,'' \emph{IEEE Signal Process. Lett.}, vol.~20, no.~3, pp. 257--260, Mar. 2013.

\bibitem{gao2014quantizer}
F.~Gao, L.~Guo, H.~Li, J.~Liu, and J.~Fang, ``Quantizer design for distributed {GLRT} detection of weak signal in wireless sensor networks,'' \emph{{IEEE} Trans. Wireless Commun.}, vol.~14, no.~4, pp. 2032--2042, 2014.

\bibitem{niu2023designing}
R.~Niu and P.~K. Varshney, ``Designing local quantizers for distributed detection of a source with unknown location,'' \emph{IEEE Commun. Lett.}, vol.~27, no.~11, pp. 3113--3117, 2023.

\bibitem{Ciuonzo2013b}
D.~Ciuonzo, G.~Papa, G.~Romano, {P. Salvo Rossi}, and P.~Willett, ``One-bit decentralized detection with a {Rao} test for multisensor fusion,'' \emph{IEEE Signal Process. Lett.}, vol.~20, no.~9, pp. 861--864, 2013.

\bibitem{ciuonzo2021IotJ}
D.~Ciuonzo, P.~Salvo~Rossi, and P.~K. Varshney, ``Distributed detection in wireless sensor networks under multiplicative fading via generalized score tests,'' \emph{IEEE Internet Things J.}, vol.~8, no.~11, pp. 9059--9071, 2021.

\bibitem{cheng2019multibit}
X.~Cheng, D.~Ciuonzo, and {P. Salvo Rossi}, ``Multibit decentralized detection through fusing smart and dumb sensors based on {Rao} test,'' \emph{IEEE Trans. Aerosp. Electron. Syst.}, vol.~56, no.~2, pp. 1391--1405, 2019.

\bibitem{yang2023hybrid}
S.~Yang, Y.~Lai, A.~Jakobsson, and W.~Yi, ``Hybrid quantized signal detection with a bandwidth-constrained distributed radar system,'' \emph{{IEEE} Trans. Aerosp. Electron. Syst.}, vol.~59, no.~6, pp. 7835--7850, 2023.

\bibitem{mao2024multi}
L.~Mao, S.~Yan, Z.~Sui, and H.~Li, ``Multi-bit distributed detection of sparse stochastic signals over error-prone reporting channels,'' \emph{IEEE Trans. on Signal and Information Processing over Networks}, vol.~10, pp. 881--893, 2024.

\bibitem{Ciuonzo2012}
D.~Ciuonzo, G.~Romano, and {P. Salvo Rossi}, ``Channel-aware decision fusion in distributed {MIMO} wireless sensor networks: Decode-and-fuse vs. decode-then-fuse,'' \emph{IEEE Trans. Wireless Commun.}, vol.~11, no.~8, pp. 2976--2985, Aug. 2012.

\bibitem{Chen2004}
B.~Chen, R.~Jiang, T.~Kasetkasem, and P.~K. Varshney, ``Channel aware decision fusion in wireless sensor networks,'' \emph{IEEE Trans. Signal Process.}, vol.~52, no.~12, pp. 3454--3458, Dec. 2004.

\bibitem{al2019decision}
M.~A. Al-Jarrah, M.~A. Yaseen, A.~Al-Dweik, O.~A. Dobre, and E.~Alsusa, ``Decision fusion for {IoT}-based wireless sensor networks,'' \emph{IEEE Internet Things J.}, vol.~7, no.~2, pp. 1313--1326, 2019.

\bibitem{ciuonzo2020SIRS}
D.~Ciuonzo and P.~Salvo~Rossi, ``Channel-aware decision fusion with {Rao} test for multisensor fusion,'' in \emph{Int. Symp. on Signal Processing and Intelligent Recognition Systems (SIRS)}, 2020, pp. 267--277.

\bibitem{Berger2009}
C.~R. Berger, M.~Guerriero, S.~Zhou, and P.~K. Willett, ``{PAC} vs. {MAC} for decentralized detection using noncoherent modulation,'' \emph{{IEEE} Trans. Signal Process.}, vol.~57, no.~9, pp. 3562--3575, Sep. 2009.

\bibitem{deMaio2018adaptive}
A.~De~Maio, S.~Han, and D.~Orlando, ``Adaptive radar detectors based on the observed {FIM},'' \emph{IEEE Trans. Signal Process.}, vol.~66, no.~14, pp. 3838--3847, 2018.

\bibitem{stoica2004model}
P.~Stoica and Y.~Selen, ``Model-order selection: a review of information criterion rules,'' \emph{{IEEE} Signal Process. Mag.}, vol.~21, no.~4, pp. 36--47, 2004.

\bibitem{Rousseau2003}
D.~Rousseau, G.~V. Anand, and F.~Chapeau-Blondeau, ``Nonlinear estimation from quantized signals: Quantizer optimization and stochastic resonance,'' in \emph{Proc. 3rd Int. Symp. Physics in Signal and Image Processing}, 2003, pp. 89--92.

\bibitem{Gao2015}
F.~Gao, L.~Guo, H.~Li, and J.~Fang, ``One-bit quantization and distributed detection with an unknown scale parameter,'' \emph{Algorithms}, vol.~8, no.~3, pp. 621--631, 2015.

\bibitem{Ciuonzo2017}
D.~Ciuonzo and {P. Salvo Rossi}, ``Distributed detection of a non-cooperative target via generalized locally-optimum approaches,'' \emph{Information Fusion}, vol.~36, pp. 261--274, 2017.

\bibitem{Ciuonzo2015}
D.~Ciuonzo, {P. Salvo Rossi}, and S.~Dey, ``Massive {MIMO} channel-aware decision fusion,'' \emph{{IEEE} Trans. Signal Process.}, vol.~63, no.~3, pp. 604--619, Feb. 2015.

\end{thebibliography}

\end{document}